\documentclass[reprint, superscriptaddress, amsmath, amssymb, aps, pra, a4paper]{revtex4-2}
\usepackage{graphicx}
\usepackage{dcolumn}
\usepackage{hyperref}
\usepackage{multirow}
\usepackage{comment}
\usepackage{xcolor}
\usepackage{amsmath}
\usepackage{mathrsfs}
\usepackage{tabularx}
\usepackage{natbib}
\usepackage{physics}
\usepackage{float}

\usepackage{soul}

\begin{document}

\title{Free Space Continuous Variable Quantum Key Distribution with Discrete Phases}

\author{Anju Rani}
\email{anju@prl.res.in}
\affiliation{Quantum Science and Technology Laboratory, Physical Research Laboratory, Ahmedabad, India 380009.}
\affiliation{Indian Institute of Technology, Gandhinagar, India 382355.}

\author{Pooja Chandravanshi}
\affiliation{Quantum Science and Technology Laboratory, Physical Research Laboratory, Ahmedabad, India 380009.}

\author{Jayanth Ramakrishnan}
\affiliation{Quantum Science and Technology Laboratory, Physical Research Laboratory, Ahmedabad, India 380009.}

\author{Pravin Vaity}
\affiliation{Quantum Science and Technology Laboratory, Physical Research Laboratory, Ahmedabad, India 380009.}

\author{P. Madhusudhan}
\affiliation{Femtosecond Laser Laboratory, Physical Research Laboratory, Ahmedabad, India 380009.}

\author{Tanya Sharma}
\affiliation{Quantum Science and Technology Laboratory, Physical Research Laboratory, Ahmedabad, India 380009.}
\affiliation{Indian Institute of Technology, Gandhinagar, India 382355.}

\author{Pranav Bhardwaj}
\affiliation{Femtosecond Laser Laboratory, Physical Research Laboratory, Ahmedabad, India 380009.}
\affiliation{Indian Institute of Technology, Gandhinagar, India 382355.}

\author{Ayan Biswas}
\affiliation{Quantum Science and Technology Laboratory, Physical Research Laboratory, Ahmedabad, India 380009.}

\author{R. P. Singh}
\email{rpsingh@prl.res.in}
\affiliation{Quantum Science and Technology Laboratory, Physical Research Laboratory, Ahmedabad, India 380009.}

\date{\today}

\begin{abstract}
Quantum Key Distribution (QKD) offers unconditional security in principle. Many QKD protocols have been proposed and demonstrated to ensure secure communication between two authenticated users. Continuous variable (CV) QKD offers many advantages over discrete variable (DV) QKD since it is cost-effective, compatible with current classical communication technologies, efficient even in daylight, and gives a higher secure key rate. Keeping this in view, we demonstrate a discrete modulated CVQKD protocol in the free space which is robust against polarization drift.
We also present the simulation results with a noise model to account for the channel noise and the effects of various parameter changes on the secure key rate. These simulation results help us to verify the experimental values obtained for the implemented CVQKD.
\end{abstract}

\maketitle

\section{\label{sec:Introduction}Introduction}
With the advancement in technology, the demand for secure communication has increased. In classical communication, the security relies on the complexity of the underlying mathematical algorithm and can be easily compromised once there is enough computational advancement\,\cite{shor1994algorithms}. Quantum Key Distribution (QKD)\,\cite{ bennett1984g,gisin2002quantum} provides a secure way to distribute a key between two communicating parties, Alice and Bob. QKD uses quantum states to encode the key information, and its security completely relies on the laws of quantum mechanics, making no assumptions about the adversary's technological power\,\cite{mayers2001unconditional}. The key exchange takes place through the quantum channel and is post-processed using an authenticated classical channel.

Implementing QKD over large distances enables secure quantum communication over a global scale and involves DVQKD protocols, which require encoding key information in a single quantum state\,\cite{liao2017satellite,PhysRevLett.119.200501,ribordy2000long,Sheng-KaiLiao:90302,ursin2009space,yin2020entanglement}. The practical implementation of these QKD protocols involves various challenges, one of which is the generation of deterministic single-photons. However, achieving this in experimental setups can be difficult. As a result, in prepare and measure DVQKD protocols, weak coherent pulses are often utilized as an alternative. Nonetheless, the use of weak coherent pulses increases the risk of photon number splitting attacks. On the measurement side, the single-photon detectors are expensive and are not photon number resolving, and hence, record multi-photon events that could lead to security loopholes\,\cite{Xu_2020}. On the other hand, entanglement-based DVQKD protocols\,\cite{Mishra_2022} are unconditionally secure\,\cite{Steinlechner_2012}, but the key rate obtained is very low. 

At this stage, we need to explore another class of QKD protocols, i.e., CVQKD protocols\,\cite{https://doi.org/10.1002/qute.201800011,4.RevModPhys.8621,e17096072} that might be proven to be one of the best possible candidates. CVQKD protocols use the quadratures of the electromagnetic field to encode key information\,\cite{TCRalph1999ContinuousVQ,Gills}. These protocols are compatible with well-established classical communication technologies, thus enabling us to use existing communication infrastructure with enhanced security\,\cite{pirandola2,A.Levierrer2} provided through quantum mechanics. Further, CVQKD protocols could be implemented using standard telecommunication components with a higher key rate\,\cite{Wang2018HighKR,Schrenk2017HighRateCQ} as compared to DVQKD protocols. The state preparation step requires the use of amplitude and phase modulators, and the measurement step uses balanced homodyne detectors that are already available commercially and operate at a very high rate\,\cite{1.2GHz,300MHz,highspeed}. In addition to this, homodyne detectors are cost-effective and have high quantum efficiency at telecommunication wavelength. These protocols are efficient even at room temperature and daylight since the local oscillator acts as a spectral, temporal, and spatial filter and is robust against stray light.

According to the modulation scheme, we can divide CVQKD protocols into continuous (Gaussian) modulation CVQKD and discrete modulation CVQKD. In the former case, one performs Gaussian modulation for both amplitude and phase quadratures, like GMCS or GG02 protocols\,\cite{Grosshans2003QuantumKD, PhysRevLett.88.057902, 2013}. The latter is based on the discrete modulation of the quadratures, like quadrature amplitude modulation (QAM)\,\cite{QAM/PSK}, quadrature phase sifts keying (QPSK)\cite{A.Levierrer7,Hirano4PBHD,Hirano1security,Hirano5DMCVQKD}. Gaussian modulated protocols are pretty mature with well-defined security\,\cite{pirandola2} and have been successfully demonstrated up to a distance of hundreds of km\,\cite{PhysRevLett.125.010502} in fiber, making them efficient for metropolitan area networks. However, implementing such protocols over long distances is challenging as it is difficult to maintain good reconciliation efficiency at low signal-to-noise ratio\,(SNR)\,\cite{reconciliation,PhysRevApplied.12.054013}.
Here comes the role of discrete modulated\,(DM) CVQKD. The advantage of DM-CVQKD is that it simplifies the modulation scheme and key extraction task, which is a bit complicated in Gaussian-modulated CVQKD protocols, where one extracts the key from continuous random values. DM-CVQKD protocols are remarkable for long-distance applicability even at low SNR\,\cite{Pan:22,https://doi.org/10.48550/arxiv.2211.16862}.

In this paper, we report the implementation of a free space discrete-modulated CVQKD protocol in the lab. The paper is structured as follows. In Sec.\,\ref{sec:Theory}, the theoretical background for the protocol is discussed, and a noise model is presented to account for the channel noise. The imperfections present in the experiment are also simulated, and the simulated results are discussed. In Sec.\,\ref{sec:Experimental Setup}, the experimental setup for the four-state discrete modulation CVQKD is presented. Sec.\,\ref{sec:RnD} shows the experimental results, and we end up with concluding remarks in Sec.\,\ref{sec:Conclusion}.

\section{\label{sec:Theory}Theory and simulation}
In this Section, we discuss the theoretical aspects of the protocol implemented and present the details of the simulation performed. Further, we describe imperfections in the experimental implementation and provide models to simulate them. We end the Section with some remarks on the security of the protocol and present the simulated results. 
\subsection{Protocol Execution}
The protocol implemented in this manuscript consists of the following steps.
\begin{enumerate}
    \item Alice randomly selects from the four coherent states $\ket{\alpha e^{i\phi_{\mathrm{A}}}}$, where $\phi_{\mathrm{A}}$ is chosen from $0, \pi/2, \pi$, and $3\pi/2$ by modulating the phase of her signal. This signal is transmitted to the receiver Bob. The phases 0 and $\pi$ correspond to encoding the bit in the $\hat{q}$ basis, and $\pi/2$ and $3\pi/2$ correspond to the $\hat{p}$ basis respectively. Here, $|\alpha|^2$ is the mean photon number of the signal.
    \item Bob performs homodyne detection\,\cite{Ulf} on the received signal and randomly decides to measure the $\hat{q}$ quadrature or the $\hat{p}$ quadrature by modulating the phase of the local oscillator\,(LO), choosing $\phi_{\mathrm{B}}$ as 0 or $\pi/2$ respectively.
    \item After the exchange of signals, Alice discloses the basis in which the bit was encoded, and Bob discloses the basis in which the signal was measured. They retain the pulses for which the encoding and the measuring basis match. This process is called sifting.
    \item The quadrature probability distributions for the measurements made by Bob for various $\phi = \phi_{\mathrm{A}}-\phi_{\mathrm{B}}$ are Gaussians centered at $\pm \alpha$ for $\phi = 0 \ \rm{and} \ \pi $ respectively and at 0 for $\phi = \pi/2 \ \rm{and} \ 3\pi/2$. The probability distributions for $\phi = \pi/2 \ \rm{and} \ 3\pi/2$ are indistinguishable and hence do not contribute to the key.
    \item The measured values for $\phi = 0 \ \rm{and} \ \pi$ contribute to the key. Since in homodyne detection, the measured output values are continuous, Bob assigns a threshold $x_0$ to the sifted signals for postselection and assigns his bit value as 
    \begin{eqnarray}
        \text{bit value} = \begin{cases} 1 &\ x_{\phi}>x_0 \\ 0  &\ x_{\phi}<-x_0\\ \text{inconclusive} & -x_0<x_{\phi}<x_0. \end{cases}
    \end{eqnarray}
    \item Alice assigns her bit value as 1 for $\phi_\mathrm{A} = 0 \ \text{and} \ \pi/2$ and 0 for $\phi_\mathrm{A} = \pi \ \text{and} \ 3\pi/2$.
    \item Alice and Bob disclose a fraction of their raw key in order to perform parameter estimation and mutual information to get the final secret key.
\end{enumerate}  

In order to understand the limitations of the carried out laboratory demonstration, a simulation of the DM-CVQKD protocol was performed. 

\subsection{\label{sec: Noise model} Noise model}
\begin{figure}
    \includegraphics[scale=0.4]{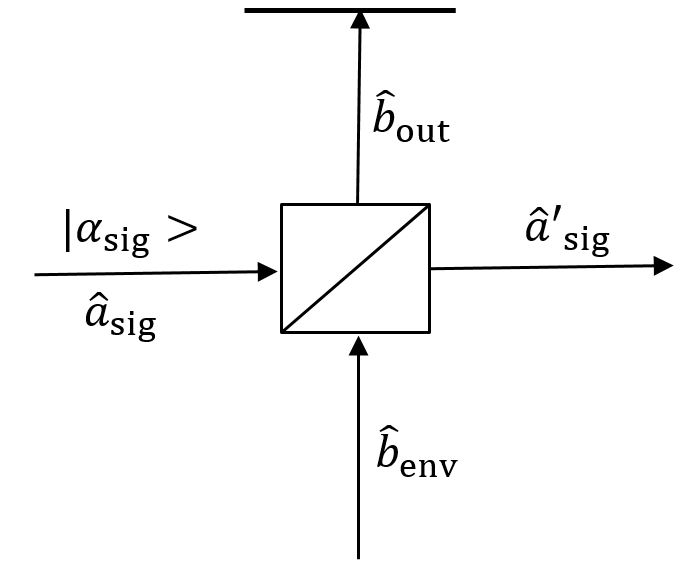}
    \caption{Theoretical model of the channel transmittance and noise included in the simulation. The beam splitter has a transmittance $T \leq 1$ and couples the quantum state  $\ket{\alpha}_{\mathrm{sig}}$ with the environment and hence introduces excess noise in input state. Here $\hat{a}_{\mathrm{sig}}$ \& $\hat{b}_{\mathrm{env}}$ represent the input field operators of signal and the environment respectively and $\hat{a}^{\prime}_{\mathrm{sig}}$ \&
    $\hat{b}_{\mathrm{out}}$ denote the output field operators after interaction at the BS.}
    \label{fig:noise_fig}
\end{figure}

 One of the major roadblocks in the implementation of quantum information protocols is the presence of noise and attenuation, which is unavoidable due to interactions of the quantum system with the environment. The state that Alice prepares is sent to Bob via a quantum channel which in reality can either be a fiber optic or a free space. The propagation of this state through the quantum channel alters the state at the output, which in turn affects Bob's measurement and introduces errors in the generated key. The effect of the transmission losses and the channel noise on the transmitted state can be evaluated by considering a model as shown in Fig. \ref{fig:noise_fig}.

A fictitious beam splitter of transmittance $T<1$ is inserted into the quantum channel separating Alice and Bob. The beam splitter couples the quantum state to the environment, which introduces noise in the state. The transmittance $T$ models the attenuation of the signal in the quantum channel.
The density matrix for the ensemble of states shared by Alice can be written as 
\begin{equation}\label{eq:state}
    \hat{\rho}_{\mathrm{sig}} = \frac{1}{4}\left(\ket{\alpha}\bra{\alpha}+\ket{-\alpha}\bra{-\alpha}+\ket{i\alpha}\bra{i\alpha}+\ket{-i\alpha}\bra{-i\alpha}\right).
\end{equation}
The effect of the channel can be evaluated by using the covariance matrix formalism\,\cite{4.RevModPhys.8621}. The covariance matrix for the state in Eq.\,\eqref{eq:state} is evaluated as
\begin{equation}
    V = \begin{pmatrix}
        \frac{|\alpha|^2}{2} + \frac{1}{4} & 0 \\
        0 & \frac{|\alpha|^2}{2} + \frac{1}{4}
    \end{pmatrix}.
\end{equation}
Here, $V_{\mathrm{mod}} =\frac{|\alpha|^2}{2}$ is Alice's modulation variance. The covariance matrix after propagation through the channel can be evaluated as
\begin{equation}
    V_{\mathrm{Bob}} = \begin{pmatrix}
        T\frac{|\alpha|^2}{2} + \frac{1}{4} + \xi_{\mathrm{ch}} & 0 \\
        0 & T\frac{|\alpha|^2}{2} + \frac{1}{4} + \xi_{\mathrm{ch}}
    \end{pmatrix},
\end{equation}
where $\xi_{\mathrm{ch}}$ is the noise added to the signal due to transmission in the channel. 

Similarly, an imperfect homodyne detection at the receiver end can also be modeled using a beam splitter with transmittance $\eta$, which denotes the detection efficiency and noise $\xi_{\mathrm{ele}}$, which models the electronic noise in shot noise units. The final covariance matrix for Alice and Bob's data will read as 
\begin{equation}
    V_{\mathrm{AB}} = \begin{pmatrix}
        \frac{|\alpha|^2}{2}\mathrm{I}_{2} & \frac{|\alpha|^2}{2}\mathrm{I}_{2} \\
        \frac{|\alpha|^2}{2}\mathrm{I}_{2} & (T\eta \frac{|\alpha|^2}{2} + \frac{1}{4} + \xi_{\mathrm{ch}} + \xi_{\mathrm{ele}})\mathrm{I}_{2} 
    \end{pmatrix},
\end{equation}
where $\mathrm{I}_2$ represents the 2x2 identity matrix.

\begin{figure}
    \includegraphics[scale=0.45]{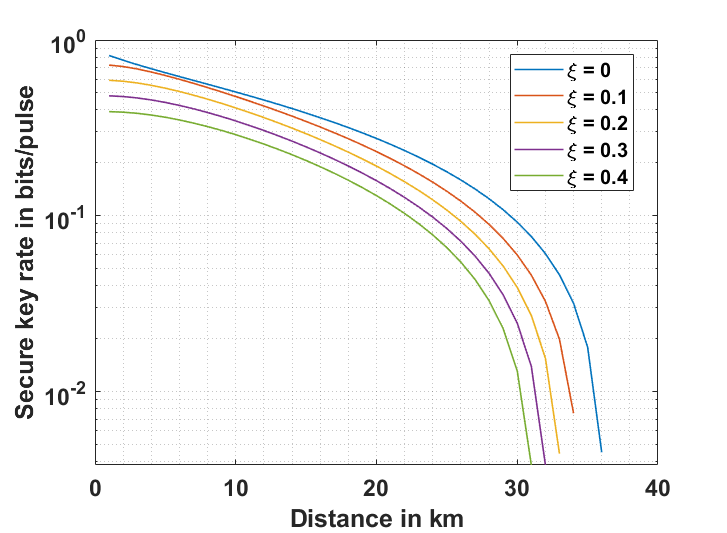}
    \caption{Plot of mutual information as a function of transmittance with different excess noises. Here, $\xi = \xi_{\mathrm{ch}} + \xi_{\mathrm{ele}}$ represents the total excess noise at Bob's end. $\xi_{\mathrm{ch}}$ denotes the noise added to the signal due to transmission in the channel \& $\xi_{\mathrm{ele}}$ denotes the electronic noise present in the detection.}
    \label{fig:mutual_information}
\end{figure}

\subsection{Mutual Information and Security}
The secret key rate \cite{Ayan} for a QKD protocol is defined by the relation,
\begin{eqnarray}
    k_{DR} &=& \beta I(A:B)- I(A:E) \  \rm{or} \label{eq:skr1}\\
    k_{RR} &=& \beta I(A:B)- I(B:E), \label{eq:skr2}
\end{eqnarray}
in the case of direct and reverse reconciliation, respectively.
 Here $I(A:B)$ is the mutual information shared between Alice and Bob, and  $I(A:E)$ or $I(B:E)$ is the information leakage to Eve in case of direct reconciliation or reverse reconciliation. $\beta$ is the reconciliation efficiency. 

For discrete modulated CVQKD under consideration, we have evaluated the mutual information between Alice and Bob by the relation,
\begin{eqnarray}
    I_{\mathrm{AB}} &=& \frac{(q_{1}+q_{2})}{2}+\frac{q_{1}}{2}\log_2(\frac{q_{1}}{(q_{1}+q_{2})})+ \nonumber \\
    &&\frac{q_{2}}{2}\log_2(\frac{q_{2}}{(q_{1}+q_{2})}),
\end{eqnarray}
where,
\begin{eqnarray}
    q_{1} &=& \mathrm{erfc}\left(\frac{(x_{0}-\sqrt{T}\alpha)}{\sqrt{2(\frac{1}{4} + \xi_{\mathrm{ch}} + \xi_{\mathrm{ele}})}}\right) \ \rm{and}\\
    q_{2} &=& \mathrm{erfc}\left(\frac{(x_{0}+\sqrt{T}\alpha)}{\sqrt{2(\frac{1}{4} + \xi_{\mathrm{ch}} + \xi_{\mathrm{ele}})}}\right).
\end{eqnarray}

In Fig.\,\ref{fig:mutual_information}, we plot the secret key rate achieved by the protocol for the case of a simple beam splitter attack by Eve. In this attack, Eve replaces the channel with a beam splitter of similar transmittance and a perfectly transmitting channel. Eve splits the signal on the beam splitter and keeps a part of the signal for measurement. The transformation on the state can be seen as
\begin{eqnarray}
    \ket{\alpha}_{\mathrm{B}}\ket{0}_{\mathrm{E}} \rightarrow \ket{\sqrt{T}\alpha}_{\mathrm{B}}\ket{\sqrt{1-T}\alpha}_{\mathrm{E}},
\end{eqnarray}
where $T$ is the transmittance of the channel, and the subscripts denote the person receiving the state. Eve then waits for the basis announcement and measures her state in the correct basis. Depending on the measurement result Eve makes a guess on the state sent by Alice. If her measured quadrature value is positive she makes a guess of Alice's bit as 1 otherwise as 0. The mutual information between Eve and Bob, I(B:E), can be evaluated and the secret key rate can be given as in Eq.\,\eqref{eq:skr2}. We have evaluated the mutual information between Bob and Eve for this particular beam splitter attack numerically, and the final secret key rate is as shown in Fig.\,\ref{fig:mutual_information}. The secret key rate has been evaluated assuming the protocol is implemented with transmittance known as a function of distance. It is seen from Fig.\,\ref{fig:mutual_information} that for experimentally relevant values of excess noise, the protocol achieves a positive key rate even up to a distance of 35 km.  

\subsection{Simulation Results}
In this Section, we have presented the simulation results obtained from our study. The results would help in a better understanding of the experimental setup and optimization of the experimental parameters.
\begin{figure}
    \includegraphics[scale = 0.4]{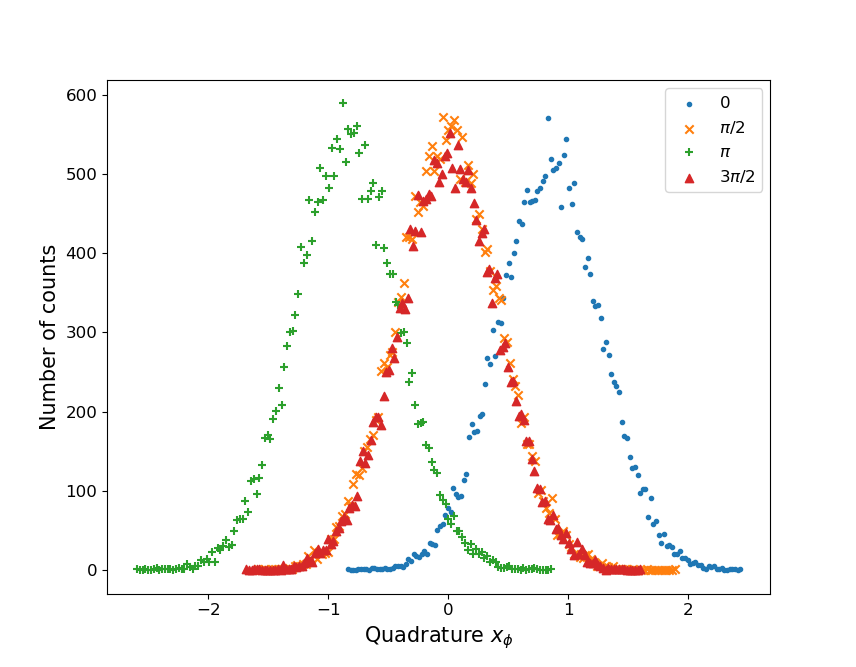}
    \caption{The simulated probability distribution of the measured homodyne output $\hat{x}_{\phi}$ corresponding to $\phi = 0, \pi/2, \pi, 3\pi/2$. Here, the mean photon number of the signal is 1. The channel transmittance $T$ was taken to be $0.9$ (under lab conditions) and the excess noise was taken to be $0.02$. The probability distributions corresponding to $\phi = \pi/2$ and $\phi = 3\pi/2$ are indistinguishable; hence, the corresponding measurements are discarded.}
    \label{fig:probdist}
\end{figure}

\begin{figure}
    \includegraphics[scale=0.425]{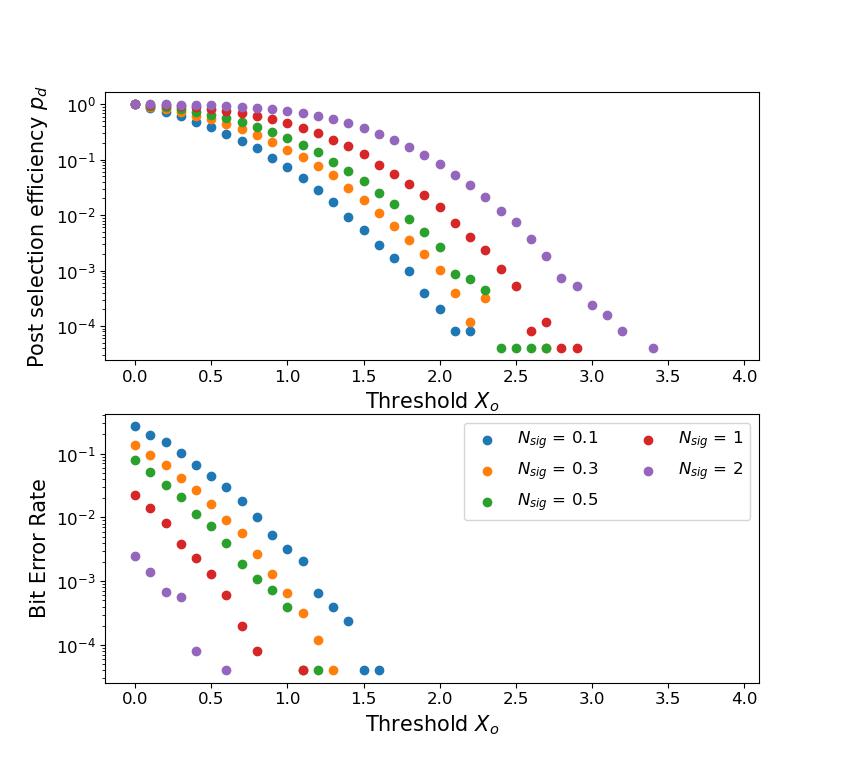}
    \caption{Plot of post-selection efficiency (top) and bit error rate (bottom) as a function of the threshold for various average photon number in the signal. The channel transmittance $T$ was taken to be $0.9$ (under lab conditions) and the excess noise was taken to be 0.02. It can be readily seen from the above graphs that on increasing the threshold $x_{\mathrm{0}}$, the bit error rate decreases; however, it also results in a decreasing post-selection efficiency which results in a lower key rate.}
    \label{fig:errors}
\end{figure}
For simulation, the channel transmittance $T$, and the excess noise were considered as $0.9$ (under lab conditions) and $0.02$, respectively. Also, the signal was taken to be a weak coherent state with an average of $1$ photon per pulse. Fig.\,\ref{fig:probdist} depicts the probability distribution of the values measured by Bob after both have disclosed their phases. It can be seen that the probability distributions for $\phi = 90^{\circ}$ and $\phi = 270^{\circ}$ are indistinguishable from each other, and hence Alice and Bob discard those measurements. The variance of the probability distribution differs from $1/4$ due to the presence of excess noise in the protocol. The mean of the probability distribution corresponding to $\phi = 0^{\circ}$ and $\phi = 180^{\circ}$ differs from $\pm 1$ due to attenuation in the channel and is given by $\pm \sqrt{T}$. Fig.\,\ref{fig:errors} depicts the post-selection efficiency and the quantum bit error rate\,(QBER) versus the threshold value selected for various mean photon numbers of the signal. It can be readily seen from Fig.\,\ref{fig:errors} that increasing the threshold value decreases the bit error rate and also decreases the post-selection efficiency. The trade-off gained by reducing the bit error rate is the reduction in post-selection efficiency which ultimately has an effect on the key rate. The simulation can help in optimizing the trade-off between bit error rate and post-selection efficiency by the optimal selection of the threshold value for the experiment being performed.

\section{\label{sec:Experimental Setup}Experimental Setup}
\begin{figure*}[t]
    \includegraphics[width=13cm]{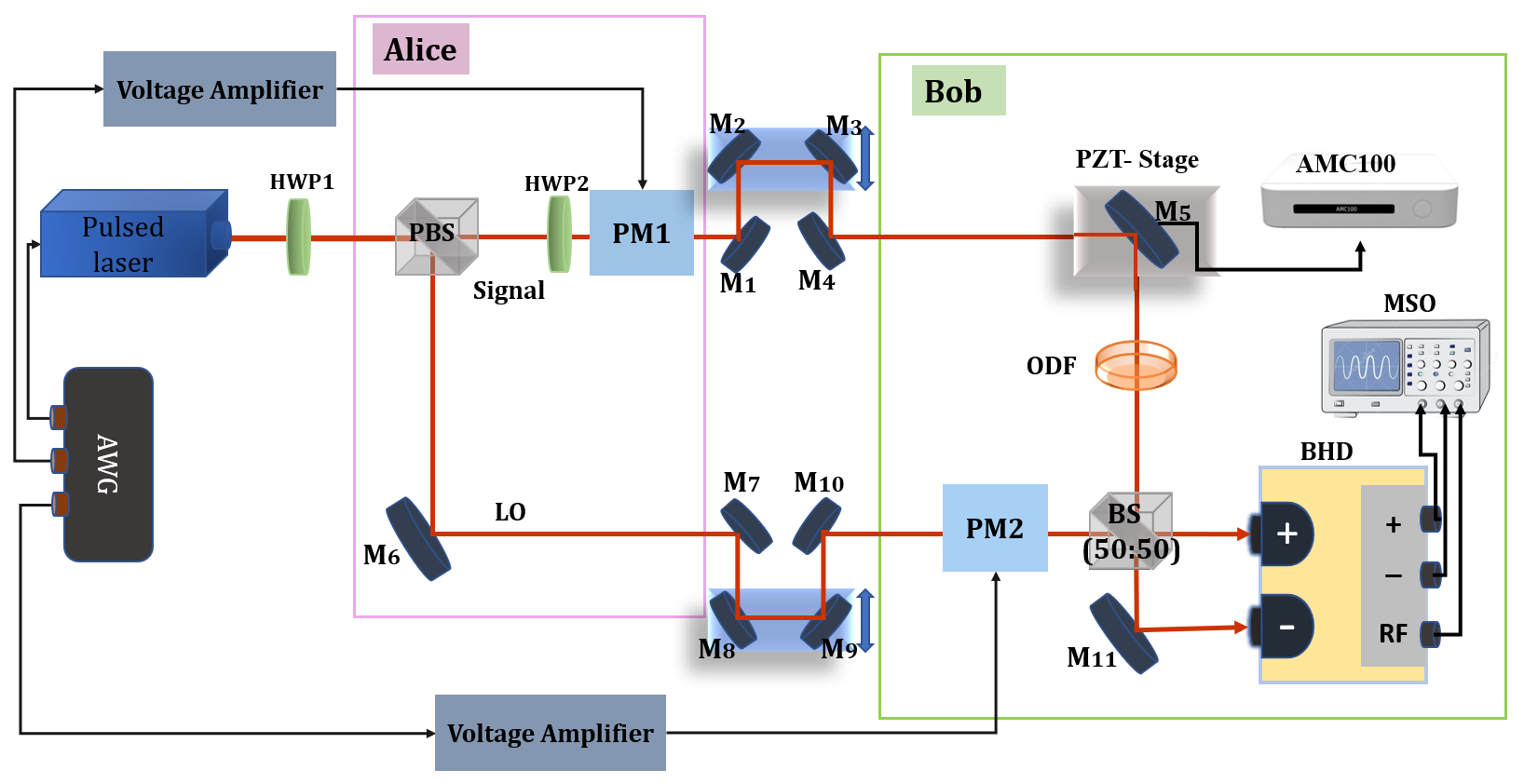} 
     \caption{Experimental scheme for free space discrete modulated CVQKD: HWP: Half Wave Plate; PBS: Polarizing Beam Splitter; PM: Electro-optic Phase Modulator; LO: Local Oscillator; M: Mirrors; PZT: Piezo Controlled Nano-positioner Stage; AMC100: Nano-positioner Controller; ODF: Optical Density Filter; BS: Beam Splitter; BHD: Balanced Homodyne Detector; MSO: Mixed Signal Oscilloscope; AWG: Arbitrary Waveform Generator.}
    \label{fig:protocol}
\end{figure*}
The experimental setup for the demonstration of discrete modulated CVQKD protocol in free space is shown in Fig.\,\ref{fig:protocol}. We have used a 780 nm pulsed laser\,(NPL79B) operating at a 1 MHz repetition rate and 30 ns pulse width. We set up a Mach Zehnder interferometer\,(MZI) for the implementation of the discrete modulated CVQKD protocol. The beam from the laser splits at a PBS into two arms of the interferometer. One arm is the signal, and the other is the local oscillator (LO). Alice controls the signal arm, whereas the LO arm is a part of Bob's detection system. We have used electro-optic phase modulators\,(EO-PM-NR-C1) to modulate the phase of Alice and Bob's signals. 

We used a high-speed AWG\,(Tektronix AWG5200) to drive a high-voltage amplifier\,(Thorlabs HVA200) which in turn drives the PM. Both signal and LO arms include four mirror alignments ($\mathrm{M_2, M_3}$ and $\mathrm{M_8, M_9}$ are placed on translation stages) to adjust the delay between them. Before using the PM, the interferometer is calibrated so as to have zero phase difference between the arms. To do this, the mirror $M_5$ is placed on a PZT-stage\,(Attocube, ECSx3080) controlled by an AMC100 controller for a fine scan of the interferometer phase. Homodyne detection is performed at the final BS. The detection system includes a balanced homodyne detector, BHD\,(Thorlab's PDB435A, DC-350 MHz), which measures the subtracted photo-current falling on the two detectors. A mixed signal oscilloscope, MSO\,(Tektronix 6-series), is used to record the output signal of BHD.

\subsection{Alice}
\label{sec:Alice} 
One arm of the interferometer i.e. the signal arm, is controlled by Alice. The phase modulator PM1 is used to encode the four-phase values for Alice i.e., ${0, \pi/2,\pi}$ and ${3\pi/2}$. The half voltage, $V_{\pi}$ of PM is 170 V. An optical density filter (ODF) with $OD = 4$ is placed in the signal arm to reduce the signal intensity. Using the combination of HWP1 and ODF,  we can control the mean photon number of the signal.

\subsection{Bob}
\label{sec:Bob}
The other arm of the interferometer, which is the LO arm, is controlled by Bob. The power of the LO is varied using the HWP1 placed before the PBS. PM2 selects the $\hat{q}$-quadrature and $\hat{p}$-quadrature values corresponding to ${0}$, and ${\pi/2}$. The mirror $\mathrm{M_5}$ is placed on a piezo nano-positioner stage to fine tune the path delay between the signal and LO arms. Bob performs homodyne detection at the final BS of the interferometer.  

\subsection{Data Acquisition}
\label{sec:Data_A} 
The phase modulation at both Alice's and Bob's ends is performed at a rate of 1 MHz. The subtracted output signal from the BHD is saved using an MSO. We have saved $8.1\text{x}10^{4}$ pulses in a single acquisition. Once sufficient data has been recorded, postprocessing is performed. We integrate the individual pulses over their respective pulse duration. Each integrated value corresponds to one quadrature value at that particular phase. We then perform sifting, and the raw key is generated. The raw key is further processed, and the secure key is obtained. Error correction and privacy amplification are performed using LDPC codes and Toeplitz hashing, respectively.

\section{\label{sec:RnD}Results and discussion}
In this Section, we present the results of our experimental implementation of the protocol.
The initial step in implementing the DM-CVQKD protocol is balancing the measurement setup\,(not shown in Fig\,\ref{fig:protocol}) and measuring the shot noise variance of the laser source\,\cite{PhysRevApplied.13.024058,PhysRevA.91.022307}. To perform the intial calibration, the signal arm is blocked, and the difference signal is measured as a function of the LO power. This measurement is used to find out the shot noise and define the shot noise unit\,(SNU) for the experiment. Once the initial calibration is done, the power of the LO is fixed at 0.25 mW. The electronic noise-to-shot noise\,(electronic-to-shot noise ratio) clearance is found to be 3.7\%. We then proceed with the implementation of the discrete modulation CV-QKD protocol.

\begin{figure}
    \includegraphics[width=01\linewidth]{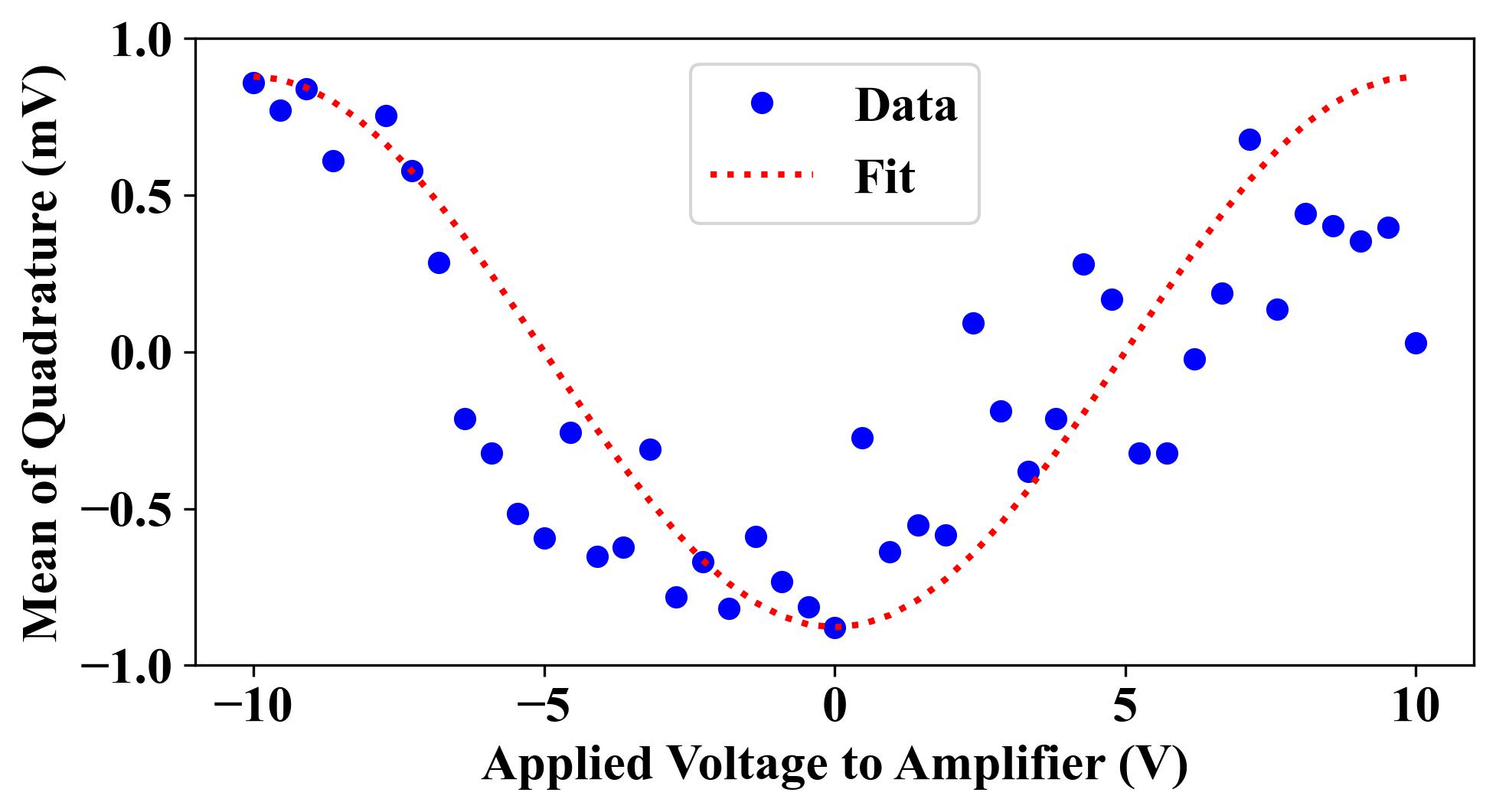}
    \caption{The variation of the mean $\hat{q}$ quadrature value of the signal as a function of the applied voltage to the PM. The voltage being applied to the PM is amplified using a voltage amplifier with a gain of -20X.}
    \label{fig:mean_volt}
\end{figure}

The interferometer is calibrated to achieve zero path difference between the arms. The condition for constructive and destructive interference is achieved with a visibility of 98\%. The signal is attenuated by using an optical density filter\,(ODF) of OD = 4 with an input power of 60 $\mu W$ before ODF. The delay introduced by the ODF is compensated by scanning the translation stage and PZT stage. The phase of the signal is then varied from $0$ to $2\pi$ by applying an appropriate voltage to the PM and the $\hat{q}$ quadrature is measured using homodyne detection. For each applied voltage, 2000 pulses are saved, and the mean of the integrated values for the pulses are plotted as a function of the applied voltage as shown in  Fig\,\ref{fig:mean_volt}.  
The fluctuation in the data is due to the inherent phase instability of the Mach-Zehnder interferometer.
\par
\begin{figure}
    \includegraphics[width=01\linewidth]{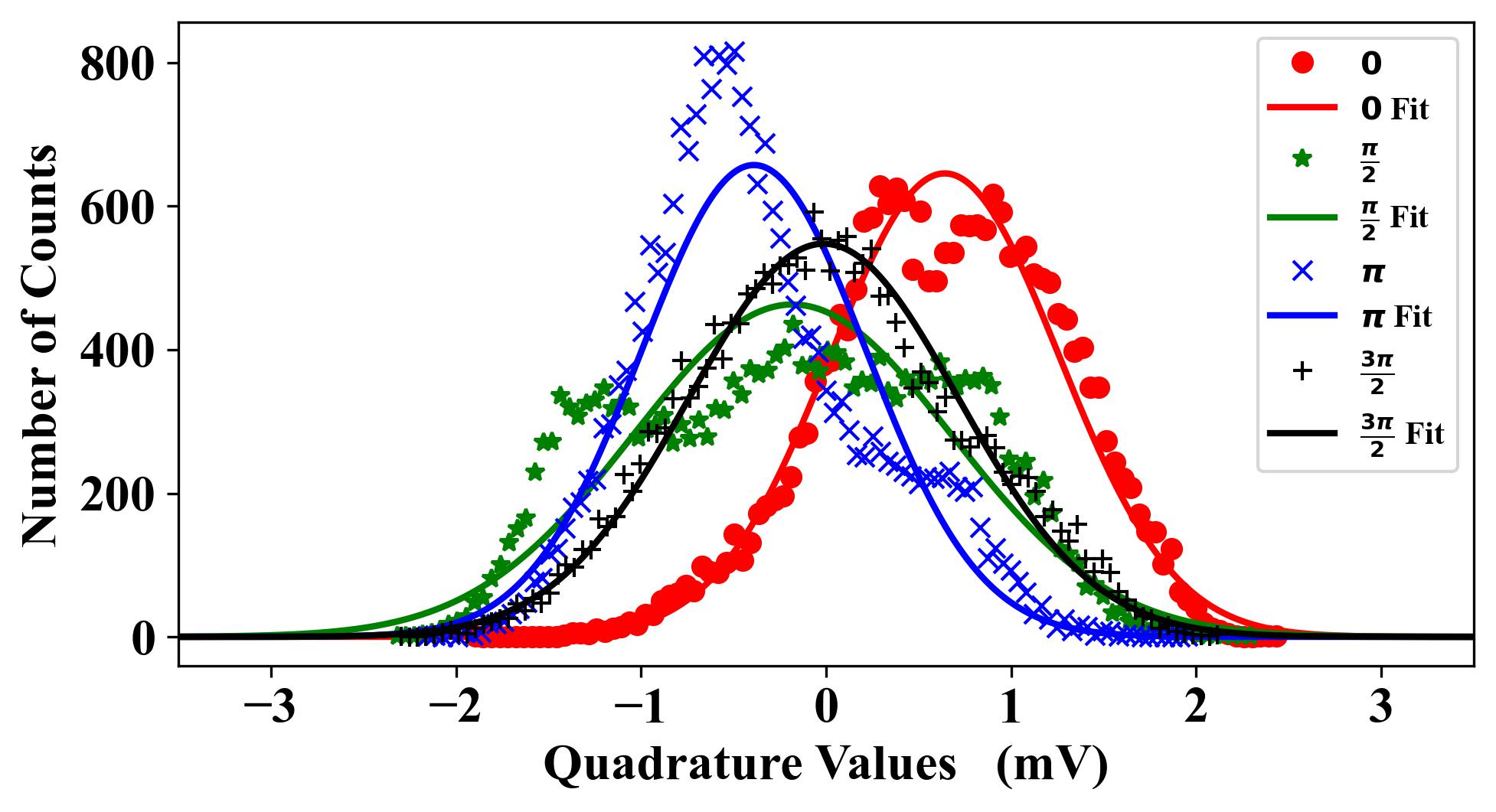}
    \caption{Probability distributions of the homodyne detected signal for the four relative phases between signal and LO. The points represent the experimental data, and the curves represent the best fit.}
    \label{fig:Dis}
\end{figure}

 A proof of principle experimental demonstration of free space DM-CVQKD has been performed. The voltages fed to both Alice and Bob's PM are generated randomly using an AWG, shown in Fig.\,\ref{fig:protocol}. A single acquisition in the MSO contains $8.1\text{x}10^{4}$ pulses. In order to retrieve the quadrature values from the signal, pulses are integrated over the respective time window. We do the basis sifting for Alice and Bob's data. The sifted key has a length of $4\text{x}10^{4}$ bits. The probability distributions of the quadrature values corresponding to relative phases are plotted in Fig\,\ref{fig:Dis}. The threshold value $x_{0}$ chosen for the experiment is 0.

We do the further post-processing of the data. For our laboratory experiments, the channel transmittance, $T = 0.95$, and detector efficiency $\eta = 0.76$ are observed. We calculated the mutual information between Alice and Bob, and finally, the secure key rate is extracted. The experimental parameters are shown in Table\,\ref{table:table2}.

\begin{table}
    \centering
    \begin{tabular}{|c|ccc|}
        \hline
        Parameters & \multicolumn{1}{c|}{Values} 
        \\ \hline
        Signal processed& \multicolumn{1}{c|}{$8.1\text{x}10^{4}$ pulses}
        \\ \hline
        Sifted bits & \multicolumn{1}{c|}{$4\text{x}10^{4}$ bits} 
        \\ \hline
        PSE & \multicolumn{1}{c|}{$3.2\text{x}10^{4}$ bits} 
        \\ \hline
        QBER & \multicolumn{1}{c|}{5\%}
        \\ \hline
        Secure key rate & \multicolumn{1}{c|}{0.35 (bit/pulse)}
        \\ \hline
    \end{tabular}
    \caption{The experimental results for the executed protocol for a single acquisition window. Here, PSE is the Post Selection Efficiency and QBER is the Quantum Bit Error Rate.}
    \label{table:table2}
\end{table} 
 
While performing CVQKD experiments, the very important parameter is the phase fluctuation of the Mach-Zehnder interferometer that affects the key rate. To account for these fluctuations, we are working on the phase stabilization of the MZI. To maximize the key rate, we will consider the noises introduced due to various sources present in the experiment in the near future.

\section{\label{sec:Conclusion}Conclusion}
We have performed a prototype tabletop experiment of the discrete modulated CVQKD protocol and have used the results to extract a secure key. We have also performed a simulation with a realistic noise model encountered in field demonstrations. The trade-off between the secure key rate and the bit error rate is illustrated using the results of the simulation. These studies assist in surveying the feasibility of continuous variable-based QKD protocols for ground as well as satellite-based communication systems. Conclusively, we can say that continuous variable-based QKD protocols can be perceived as the next frontier in the field of secure communication, be it fiber, free space, or satellite-to-ground communication.

\section*{\label{sec:Acknowledgments}Acknowledgments}
We thank Dr. Rajesh Kumar Kushawaha for providing the resources required for the experiment. We thank Dr. Rupesh Kumar, Dayanand Mishra, Jaya Krishna Meka, and QST lab members for their valuable input. The authors acknowledge the financial support from DST through the QuEST program.

\section*{\label{sec:Disclosure}Disclosures}
The authors declare no conflicts of interest related to this article.

\appendix
\section{Noise Model}
\title{Appendix A}

Consider the field operator $\hat{a}_{\mathrm{sig}}$ of the signal and $\hat{b}_{\mathrm{env}}$ of the environment. The signal is in a coherent state which is given by $\ket{\alpha}_{\mathrm{sig}}$ 

The action of the beam splitter on the field operators are given by
\begin{equation}
    \begin{pmatrix}
    \hat{a}^{\prime}_{\mathrm{sig}} \\
    \hat{b}_{\mathrm{out}}
    \end{pmatrix} = 
    \begin{pmatrix}
    \sqrt{T} & \sqrt{1-T} \\
    -\sqrt{1-T} & \sqrt{T}
    \end{pmatrix}
    \begin{pmatrix}
    \hat{a}_{\mathrm{sig}} \\
    \hat{b}_{\mathrm{env}}
    \end{pmatrix}.\label{eq:bsmat}
\end{equation}
The mode represented by the field operator $\hat{a}_{\mathrm{out}}$ is received by Bob, who performs a measurement on the corresponding quantum state. Since we are dealing with Gaussian states and the noise model represents a Gaussian transformation on the modes, we can utilize the elegant variance matrix formalism to understand the effect of the quantum channel on the state. The covariance matrix of a single-mode Gaussian state is given by
\begin{equation} \label{eq:covmat}
    V_{ij} = \frac{1}{2}\left\langle \acomm{\hat{x}_i}{\hat{x}_j}\right\rangle - \langle \hat{x}_i\rangle \langle \hat{x}_j\rangle,
\end{equation}
where $\hat{\textbf{x}} = [\hat{q}, \hat{p}]^{\mathrm{T}}$ are the quadrature operators of the signal mode given by $\hat{q} =\frac{1}{2}( \hat{a}_{\mathrm{sig}} + \hat{a}^{\dagger}_{\mathrm{sig}}) $ and $\hat{p} = \frac{i}{2}(\hat{a}^{\dagger}_{\mathrm{sig}} - \hat{a}_{\mathrm{sig}})$,
and $\acomm{\hat{A}}{\hat{B}} = \hat{A}\hat{B} + \hat{B}\hat{A}$ denotes the anti-commutator of operators $\hat{A}$ and $\hat{B}$. For the example of a coherent state the covariance matrix reduces to 
\begin{equation}
    V = \frac{1}{4}
    \begin{pmatrix}
        1 & 0 \\
        0 & 1
    \end{pmatrix}.
\end{equation}
Using Eq. \eqref{eq:bsmat}, the quadrature operators of the output signal can be written as 
\begin{eqnarray}
    \hat{q}^{\prime}_{\mathrm{sig}} &=& \sqrt{T}\hat{q}_{\mathrm{sig}}+\sqrt{1-T}\hat{q}_{\mathrm{env}} \quad {\rm and} \\
    \hat{p}^{\prime}_{\mathrm{sig}} &=& \sqrt{T}\hat{p}_{\mathrm{sig}}+\sqrt{1-T}\hat{p}_{\mathrm{env}}.
\end{eqnarray}
The combined covariance matrix of the signal and the environment after the action of the beam splitter is given by
\begin{equation}\label{eq:final}
    \Sigma = \mathrm{BS}\begin{pmatrix}
        \frac{1}{4} \mathrm{I}_{2} & 0_2 \\
        0_2 & N_0\mathrm{I}_{2}
    \end{pmatrix}\mathrm{BS}^T ,
\end{equation}
where $N_0$ denotes the channel noise and the matrix $\mathrm{BS}$ is defined as
\begin{equation}\label{eq:bsmatrix}
    \mathrm{BS} = \begin{pmatrix}
        \sqrt{T}\mathrm{I}_{2} & \sqrt{1-T}\mathrm{I}_{2} \\
        -\sqrt{1-T}\mathrm{I}_{2} & \sqrt{T}\mathrm{I}_{2}
    \end{pmatrix}.
\end{equation}
Evaluating the expression given in Eq.\,\ref{eq:final}, the covariance matrix of the singal reaching Bob is given by
\begin{equation}
    V_{\mathrm{Bob}} = \begin{pmatrix}
        T\frac{|\alpha|^2}{2} + \frac{1}{4} + \xi_{\mathrm{ch}} & 0 \\
        0 & T\frac{|\alpha|^2}{2} + \frac{1}{4} + \xi_{\mathrm{ch}}
    \end{pmatrix},
\end{equation}
where $N_0 = \frac{1}{4}+(\xi_{\mathrm{ch}}/(1-T))$.

\bibliography{reference}

\end{document}